\documentclass[%
 reprint,
 amsmath,amssymb,
 aps,
]{revtex4-1}

\usepackage{graphicx}
\usepackage{dcolumn}
\usepackage{bm}
\usepackage{color}

\begin{document}

\preprint{APS/123-QED}

\title{Examination of the influence of the f$_{0}$(975) and $\phi$(1020) mesons on the surface gravitional redshift of the neutron star PSR J0348+0432}

\author{Xian-Feng Zhao$^{1,2}$}
\email{zhaopioneer.student@sina.com}
 \affiliation{%
$^{1}$School of Sciences, Southwest Petroleum University, Chengdu, 610500, China\\
$^{2}$School of Electronic and Electrical Engineering, Chuzhou University, Chuzhou, 239000, China
}%



\date{February 2, 2015}

\begin{abstract}
The effect of the mesons $f_{0}(975)$ and $\phi(1020)$ on the surface gravitional redshift of the neutron star PSR J0348+0432 is examined in the framework of the relativistic mean field theory by choosing the suitable hyperon coupling constants. We find that compared with that without considering the mesons $f_{0}(975)$ and $\phi(1020)$, the value range of the radius $R$ of the neutron star PSR J0348+0432 would be changed from a narrow range 12.964 km $\sim$ 12.364 km to a wider range 12.941 km $\sim$ 11.907 km corresponding to the observation mass M=1.97 M$_{\odot}$$\sim$2.05 M$_{\odot}$. We also find that the value range of the surface gravitational redshift $z$ of the neutron star PSR J0348+0432 changes from 0.3469 $\sim$ 0.3997 to 0.3480 $\sim$ 0.4263 corresponding to the observation mass M=1.97 M$_{\odot}$$\sim$2.05 M$_{\odot}$ as the mesons $f_{0}(975)$ and $\phi(1020)$ being considered. These mean the radius $R$ and the surface gravitational redshift $z$ all will be constrained in a wider scope as the mesons $f_{0}(975)$ and $\phi(1020)$ being considered. We also can see that the difference of the radius and the surface gravitational redshift is not so large whether the mesons $f_{0}(975)$ and $\phi(1020)$ being considered or not. This indicates that the mesons $f_{0}(975)$ and $\phi(1020)$ do not play a major role in the massive neutron star PSR J0348+0432.
\begin{description}
\item[PACS numbers]
21.65.+f, 24.10.Jv,26.60.+c,21.30.Fe  
\end{description}
\end{abstract}

\maketitle


\section{Introduction}
In 2010, Demorest et al first observed the massive neutron star PSR J1614-2230~\cite{Dem10}, which was soon theoretically studied by many researchers with various methods~\cite{{Mass12},{Masu12},{Malli12},{Whitt12},{Kapu90 91},{Jiang12},{Weiss12},{Cham12},{Kata12},{Weis12},{Bedn12}}.

Recently, a more massive neutron star PSR J0348+0432 with the mass of $2.01\pm0.04$ M$_{\odot}$ was obsearved by Antoniadis et al in 2013~\cite{Anton13}. The neutron star mass would restrict its properties, such as energy density, pressure, baryon number density and chemical potential, which depend upon the neutron star maximum mass~\cite{Lattimer10}.

Neutron stars, especially massive neutron stars are high density objects. Within it the hyperons should produce. The interactions between nucleon-nucleon or nucleon-hyperon can be represented by mesons $\sigma, \omega$ and $\rho$~\cite{Glendenning85}. But the interactions between the nucleons and hyperons can be represented by mesons $f_{0}(975)$ and $\phi(1020)$~\cite{Schaffner94}.

In the calculations of the neutron star matter with the relativistic mean field (RMF) theory, the nucleon coupling constants and the hyperon coupling constants should be determined. The theoretical results showed that the nucleon coupling constants GL85 is a better parameters to describe the properties of the neutron star matter~\cite{Glendenning85}. For the hyperon coupling constants, there are many selection methods. For example, we can select the hyperon coupling constants of mesons $\rho, \omega$ by SU(6) symmetry and those of mesons $\sigma$ by fitting the $\Lambda, \Sigma$ and $\Xi$ well depth in nuclear matter~\cite{Zhao12}. This selection methods may connect the latest progress of the nuclear physics experiments with the observation of the neutron star.

The surface gravitational redshift is closely linked to the neutron star mass and so it is also an important physical quantity~\cite{Glen97}. But the experimental data and theoretical study on the surface gravitational redshift of the neutron star PSR J0348+0432 is very few. So theoretical study on it is very necessary.

In this paper, on the hadronic basis we examine the effect of $f_{0}(975)$s and $\phi(1020)$s on the surface gravitational redshift of the massive neutron star PSR J0348+0432 in the framework of the RMF theory considering the baryon octet.

\section{The RMF theory of a neutron star}
The Lagrangian density of hadron matter containing mesons $f_{0}(975)$ and $\phi(1020)$ reads as follows~\cite{Glen97}
\begin{eqnarray}
\mathcal{L}&=&
\sum_{B}\overline{\Psi}_{B}(i\gamma_{\mu}\partial^{\mu}-{m}_{B}+g_{\sigma B}\sigma-g_{\omega B}\gamma_{\mu}\omega^{\mu}
\nonumber\\
&&-\frac{1}{2}g_{\rho B}\gamma_{\mu}\tau\cdot\rho^{\mu})\Psi_{B}+\frac{1}{2}\left(\partial_{\mu}\sigma\partial^{\mu}\sigma-m_{\sigma}^{2}\sigma^{2}\right)
\nonumber\\
&&-\frac{1}{4}\omega_{\mu \nu}\omega^{\mu \nu}+\frac{1}{2}m_{\omega}^{2}\omega_{\mu}\omega^{\mu}-\frac{1}{4}\rho_{\mu \nu}\cdot\rho^{\mu \nu}
\nonumber\\
&&+\frac{1}{2}m_{\rho}^{2}\rho_{\mu}\cdot\rho^\mu-\frac{1}{3}g_{2}\sigma^{3}-\frac{1}{4}g_{3}\sigma^{4}
\nonumber\\
&&+\sum_{\lambda=e,\mu}\overline{\Psi}_{\lambda}\left(i\gamma_{\mu}\partial^{\mu}
-m_{\lambda}\right)\Psi_{\lambda}
\nonumber\\
&&+\mathcal{L}^{YY}
.\
\end{eqnarray}
The last term represents the contribution of the mesons $f_{0}(975)$s and $\phi(1020)$s and reads
\begin{eqnarray}
\mathcal{L}^{YY}&=&\sum_{B}g_{f_{0} B}\overline{\Psi}_{B}\Psi_{B}f_{0}-\sum_{B}g_{\phi B}\overline{\Psi}_{B}\gamma_{\mu}\Psi_{B}\phi^{\mu}
\nonumber\\
&&+\frac{1}{2}\left(\partial_{\mu}f_{0}\partial^{\mu}f_{0}-m_{f_{0}}^{2}f_{0}^{2}\right)-\frac{1}{4}S_{\mu \nu}S^{\mu \nu}
+\frac{1}{2}m_{\phi}^{2}\phi_{\mu}\phi^{\mu}
.\
\end{eqnarray}
Here, $S_{\mu \nu}=\partial_{\mu}\phi_{\nu}-\partial_{\nu}\phi_{\mu}$.

Then the RMF approach is used. From the condition of $\beta$ equilibrium in neutron star matter, the chemical equilibrium is
\begin{eqnarray}
\mu_{i}=b_{i}\mu_{n}-q_{i}\mu_{e},
\end{eqnarray}
where $b_{i}$ is the baryon number of a species $i$.
\section{The parameters}
The nucleon coupling constant GL85 set is chosen in this work~\cite{Glendenning85}: the saturation density $\rho_{0}$=0.145 fm$^{-3}$, binding energy B/A=15.95 MeV, a compression modulus $K=285$ MeV, charge symmetry coefficient $a_{sym}$=36.8 MeV and the effective mass $m^{*}/m$=0.77.

We define the ratios of hyperon coupling constant to nucleon coupling constant: $x_{\sigma h}=\frac{g_{\sigma h}}{g_{\sigma}}=x_{\sigma}$, $x_{\omega h}=\frac{g_{\omega h}}{g_{\omega}}=x_{\omega}$, $x_{\rho h}=\frac{g_{\rho h}}{g_{\rho}}$, with $h$ denoting hyperons $\Lambda, \Sigma$ and $\Xi$.

We select $x_{\rho \Lambda}=0, x_{\rho \Sigma}=2, x_{\rho \Xi}=1$ according to SU(6) symmetry~\cite{Schaff96}. The experimental data of the hyperon well depth are $U_{\Lambda}^{(N)}=-30$ MeV~\cite{Batt97}, $ U_{\Sigma}^{(N)}=10\sim40$ MeV~\cite{{Kohno06},{Harada05},{Harada06},{Fried07}} and $U_{\Xi}^{(N)}=-28$ MeV~\cite{Schaff00}, respectively. We then choose $U_{\Lambda}^{(N)}=-30$ MeV, $ U_{\Sigma}^{(N)}$=+40 MeV and $U_{\Xi}^{(N)}=-28$ MeV in this work.

The calculations show that the ratio of hyperon coupling constant to nucleon coupling constant is in the range of $\sim$ 1/3 to 1~\cite{Glen91}. So we choose $x_{\sigma \Lambda}$=0.4, 0.5, 0.6, 0.7, 0.8, 0.9 at first and considering the restriction of the hyperon well depth~\cite{Glen97}

\begin{eqnarray}
U_{h}^{(N)}=m_{n}\left(\frac{m_{n}^{*}}{m_{n}}-1\right)x_{\sigma h}+\left(\frac{g_{\omega}}{m_{\omega}}\right)^{2}\rho_{0}x_{\omega h}
,\
\end{eqnarray}
the hyperon coupling constants $x_{\omega \Lambda}$ will be obtained. As $x_{\sigma \Lambda}$=0.9, $x_{\omega \Lambda}$ will be 1.0729 considering the restriction of the hyperon well depth. Therefore, we choose $x_{\sigma \Lambda}$=0.4, 0.5, 0.6, 0.7, 0.8 and correspondingly we obtain $x_{\omega \Lambda}$=0.3679, 0.5090, 0.6500, 0.7909, 0.9319, respectively.

As $x_{\sigma \Sigma}$=0.6, 0.7, 0.8, 0.9, $x_{\omega \Sigma}$ all will be greater than 1 (e.g. $x_{\sigma \Sigma}$=0.6, $x_{\omega \Sigma}$=1.1069). So we choose $x_{\sigma \Sigma}$=0.4, 0.5, correspondingly we obtain $x_{\omega \Sigma}$=0.8250, 0.9660. For the positive $ U_{\Sigma}^{(N)}$ restricting the production of the hyperon $\Sigma$~\cite{Zhao11}, so we only choose $x_{\sigma \Sigma}=0.4$, $x_{\omega \Sigma}$=0.8250 while $x_{\sigma \Sigma}=0.5$, $x_{\omega \Sigma}$=0.9660 can be deleted (see Table~\ref{tab1}).

\begin{table}
\begin{center}
\caption{The hyperon coupling constants fitted to the experimental data of the well depth, which are $U_{\Lambda}^{N}=-30$ MeV, $U_{\Sigma}^{N}=+40$ MeV and $U_{\Xi}^{N}=-28$ MeV, respectively. }
\label{tab1}
\begin{tabular}{p{1.2cm}<{\centering}p{0.5cm}<{\centering}p{1.1cm}<{\centering}p{1.1cm}<{\centering}p{1.1cm}<{\centering}p{1.1cm}<{\centering}p{1.1cm}<{\centering}
                p{0.5cm}<{\centering}p{0.5cm}<{\centering}p{1.1cm}<{\centering}}
\hline\noalign{\smallskip}
NO.&$x_{\sigma \Lambda}$ &$x_{\omega \Lambda}$&$x_{\sigma \Sigma}$  &$x_{\omega \Sigma}$ &$x_{\sigma \Xi}$     &$x_{\omega \Xi}$    \\
\hline
(1-30)&0.4               &0.3679                  &0.4               &0.8250              &0.4134               &0.4              \\
&0.5               &0.5090                  &\underline{0.5}   &\underline{0.9660}  &0.4843               &0.5                \\
&0.6               &0.6500                  &                  &                    &0.5553               &0.6              \\
&0.7               &0.7909                  &                  &                    &0.6262               &0.7              \\
&0.8               &0.9319                  &                  &                    &0.6971               &0.8              \\
&                  &                        &                  &                    &0.7681               &0.9              \\
\hline
31&  0.8           & 0.9319            &  0.4            &0.8250               &0.7752               &0.91              \\
32&  0.8           & 0.9319            &  0.4            &0.8250               &0.7893               &0.93
 \\
33&  0.8           & 0.9319            &  0.4            &0.8250               &0.8035               &0.95
 \\
34&  0.8           & 0.9319            &  0.4            &0.8250               &0.8177               &0.97              \\
35&  0.8           & 0.9319            &  0.4            &0.8250               &0.8319               &0.99              \\
\hline
36&  0.8           & 0.9319            &  0.4            &0.8250               &0.8326               &0.991              \\
37&  0.8           & 0.9319            &  0.4            &0.8250               &0.8340               &0.993
 \\
38&  0.8           & 0.9319            &  0.4            &0.8250               &0.8355               &0.995
 \\
39&  0.8           & 0.9319            &  0.4            &0.8250               &0.8369               &0.997
\\
40&  0.8           & 0.9319            &  0.4            &0.8250               &0.8383               &0.999     \\
\noalign{\smallskip}\hline
\end{tabular}
\end{center}
\end{table}

For $x_{\omega \Xi}$, we first choose $x_{\omega \Xi}$=0.4, 0.5, 0.6, 0.7, 0.8, 0.9 and the $x_{\sigma \Xi}$ is obtained through fitting to the hyperon well depth. Thus, the parameters chosen are listed in Table~\ref{tab1}.

\begin{figure}[!htp]
\begin{center}
\includegraphics[width=3.5in]{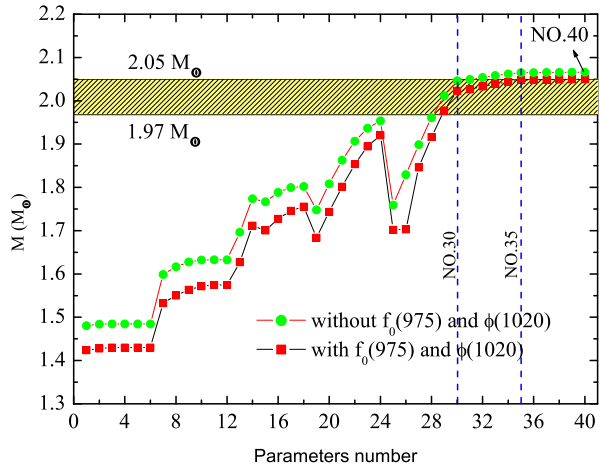}
\caption{(Color online) The mass of the neutron star as a function of the parameter numbers.}
\label{fig1}
\end{center}
\end{figure}

From the parameters chosen above, we can compose of 30 sets of suitable parameters(named as NO.1-30), for which one we calculate the mass of the neutron star (see Fig.~\ref{fig1}) through the Oppenheimer-Volkoff ( O-V ) equation~\cite{Glen97}

\begin{eqnarray}
\frac{\mathrm dp}{\mathrm dr}&=&-\frac{\left(p+\varepsilon\right)\left(M+4\pi r^{3}p\right)}{r \left(r-2M \right)}
,\\\
M&=&4\pi\int_{0}^{r}\varepsilon r^{2}\mathrm dr
.\
\end{eqnarray}

Considering the effect of the mesons $f_{0}(975)$ and $\phi(1020)$, the parameters can be chosen as follows.

The coupling constants of the mesons $\phi(1020)$ is obtained by the quark model relationships
\begin{eqnarray}
g_{\phi \Xi}=2g_{\phi \Lambda}=-2\sqrt{2}g_{\omega}/3
.\
\end{eqnarray}
For the mesons $f_{0}(975)$, we use the mass of the obtained $f_{0}(975)$ meson, but treat its couplings purely phenomenologically so as to satisfy the equation of potential depths
\begin{eqnarray}
U_{\Lambda}^{(\Xi)}\simeq=U_{\Xi}^{(\Xi)}\simeq2U_{\Lambda}^{(\Lambda)}\simeq40 MeV.
\end{eqnarray}
Thus we yield $g_{f_{0} \Lambda}/g_{\sigma}=g_{f_{0} \Sigma}/g_{\sigma}=0.69$, $g_{f_{0} \Xi}/g_{\sigma}=1.25$.

For parameters NO.1 to NO.30, considering the effect of the mesons $f_{0}(975)$ and $\phi(1020)$ the mass obtained also shown in Fig.~\ref{fig1}.

From Fig.~\ref{fig1} we see that the mass including the mesons $f_{0}(975)$ and $\phi(1020)$ are less than those not considering those two mesons. Whether considering the effect of the mesons $f_{0}(975)$ and $\phi(1020)$ or not, parameters NO.30 ($x_{\sigma \Lambda}$=0.8, $x_{\omega \Lambda}$=0.9319; $x_{\sigma \Sigma}$=0.4, $x_{\omega \Sigma}$=0.8250; $x_{\sigma \Xi}$=0.7681, $x_{\omega \Xi}$=0.9;) gives the largest mass of neutron star, which are less than 2.05 M$_{\odot}$.

In order to obtain more lager mass of the neutron star, we choose $x_{\omega \Xi}$=0.91, 0.93, 0.95, 0.97, 0.99 and we obtain $x_{\sigma \Xi}$=0.7752, 0.7893, 0.8035, 0.8177, 0.8319 by fitting to the well depth $U_{\Xi}^{(N)}$=-28 MeV. Thus we get parameters NO.31 to NO.35, respectively. We find the largest mass obtained ( by NO.35 ) also less than 2.05 M$_{\odot}$ considering the effect of the mesons $f_{0}(975)$ and $\phi(1020)$. So, by further choosing $x_{\omega \Xi}$=0.991, 0.993, 0.995, 0.997, 0.999, we obtain the corresponding  five $x_{\sigma \Xi}$s considering the constraints of the well depth $U_{\Xi}^{(N)}$=-28 MeV and we therefore get five sets of parameters named as NO.36 to NO.40. We find that the largest mass obtained by NO.40 is M=2.05 M$_{\odot}$ considering the mesons $f_{0}(975)$ and $\phi(1020)$.

In the next step, we use parameters NO.40 ($x_{\sigma \Lambda}$=0.8, $x_{\omega \Lambda}$=0.9319; $x_{\sigma \Sigma}$=0.4, $x_{\omega \Sigma}$=0.8250; $x_{\sigma \Xi}$=0.8383, $x_{\omega \Xi}$=0.999;) to study the effect of $f_{0}(975)$s and $\phi(1020)$s on the surface gravitational redshift of the massive neutron star PSR J0348+0432.

\section{The gravitational redshift of the massive neutron star PSR J0348+0432 without $f_{0}(975)$ and $\phi(1020)$ mesons}
For the neutron star PSR J0348+0432, its mass is in the range $\sim$ 1.97 M$_{\odot}$$\leq$ \emph{M} $\leq$2.05 M$_{\odot}$.

In the first case, we assume that the neutron star PSR J0348+0432 does not containing the mesons $f_{0}(975)$ and $\phi(1020)$ and parameters NO.40 is used.

The radius of PSR J0348 +0432 calculated by us is R=12.964 km corresponding to M=1.97 M $_{\odot}$ and is R=12.364 km as M=2.05 M $_{\odot}$. In this case, the radius ( R ) range of the neutron star PSR J0348+0432 should be suggested as 12.964 km $\sim$ 12.364 km by the observation M=1.97 M$_{\odot}$ $\sim$ 2.05 M$_{\odot}$.

\begin{figure}[!htp]
\begin{center}
\includegraphics[width=3.5in]{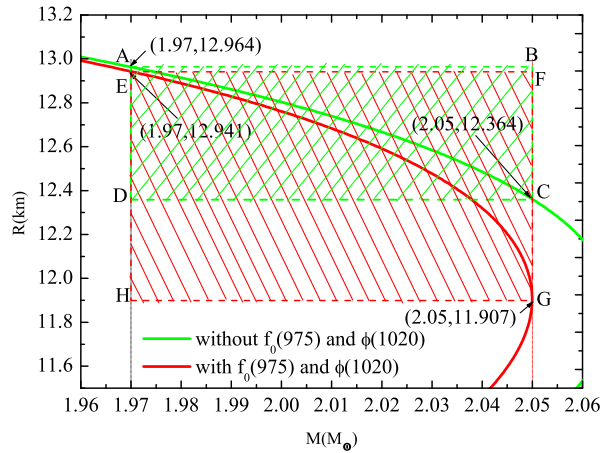}
\caption{(Color online) The radius of the neutron star as a function of the mass. Case 1 do not include the mesons $f_{0}(975)$ and $\phi(1020)$ while case 2 include them.}
\label{fig2}
\end{center}
\end{figure}

The surface gravitational redshift of a neutron star is given by~\cite{Glen97}
\begin{eqnarray}
z=(1-\frac{2M}{R})^{-1/2}-1.
\end{eqnarray}

\begin{figure}[!htp]
\begin{center}
\includegraphics[width=3.5in]{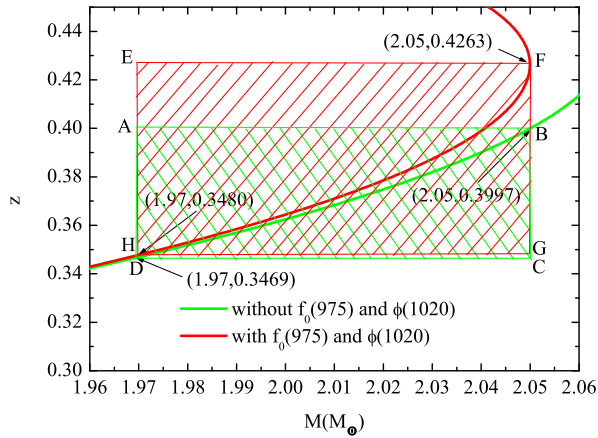}
\caption{(Color online) The radius of the neutron star as a function of the mass. Case 1 do not include the mesons $f_{0}(975)$ and $\phi(1020)$ while case 2 include them.}
\label{fig3}
\end{center}
\end{figure}

According to the M and R previously obtained, the surface gravitational redshift of the neutron star PSR J0348+0432 calculated is z=0.3469 corresponding to M=1.97 M${_\odot}$ and R=12.964 km(Fig.\ref{fig3}) and is z=0.3997 as M=2.05 M${_\odot}$ and R=12.364 km. Thus, the surface gravitational redshift of the neutron star PSR J0348+0432 should be suggested in the range 0.3469 $\sim$ 0.3997 without the mesons $f_{0}(975)$ and $\phi(1020)$ being considered.

\section{The gravitational redshift of the massive neutron star PSR J0348+0432 considering the $f_{0}(975)$ and $\phi(1020)$ mesons}
Next, we assume that the neutron star PSR J0348+0432 contains the mesons $f_{0}(975)$ and $\phi(1020)$ and parameters NO.40 is also used.

Corresponding to M=1.97 M$_{\odot}$, the calculated radius of PSR J0348 +0432 is R=12.941 km and is R=11.907 km for M=2.05 M$_{\odot}$. Here, the radius ( R ) range of the neutron star PSR J0348+0432 should be suggested as 12.941 km $\sim$ 11.907 km by the observation M=1.97 M$_{\odot}$ $\sim$ 2.05 M$_{\odot}$.

We further obtain the surface gravitational redshift of the neutron star PSR J0348+0432 according to the previously obtained mass M and radius R. The surface gravitational redshift of the neutron star PSR J0348+0432 we calculate is z=0.3480 corresponding to M=1.97 M${_\odot}$ and R=12.941 km(Fig.\ref{fig3}) and is z=0.4263 as M=2.05 M${_\odot}$ and R=11.907 km. The suggested surface gravitational redshift of the neutron star PSR J0348+0432 should be  in the range 0.3480 $\sim$ 0.4263 with the mesons $f_{0}(975)$ and $\phi(1020)$ being considered.

\section{Discussion on the gravitational redshift of the massive neutron star PSR J0348+0432 with and without the $f_{0}(975)$ and $\phi(1020)$ mesons}
From the results obtained above, we see the radius ( R ) range of the neutron star PSR J0348+0432 is 12.964 km $\sim$ 12.364 km without the $f_{0}(975)$ and $\phi(1020)$ mesons and is 12.941 km $\sim$ 11.907 km as those two mesons being considered by the observation M=1.97 M$_{\odot}$ $\sim$ 2.05 M$_{\odot}$. That is to say, considering $f_{0}(975)$ and $\phi(1020)$ mesons the radius range of the neutron star PSR J0348+0432 changes from R=12.964 km $\sim$ 12.364 km to R=12.941 km $\sim$ 11.907 km.

We also can see the surface gravitational redshift of the neutron star PSR J0348+0432 is in the range $z$=0.3469 $\sim$ 0.3997 without the $f_{0}(975)$ and $\phi(1020)$ mesons and is $z$=0.3480 $\sim$ 0.4263 as those two mesons being considered by the observation M=1.97 M$_{\odot}$ $\sim$ 2.05 M$_{\odot}$. Namely, considering $f_{0}(975)$ and $\phi(1020)$ mesons the surface gravitational redshift of the neutron star PSR J0348+0432 changes from $z$=0.3469 $\sim$ 0.3997 to $z$=0.3480 $\sim$ 0.4263.

The results above also can be seen in Table~\ref{tab2}.

\begin{table}[!htbp]
\begin{center}
\caption{The radius $R$ and the surface gravitational redshift $z$ calculated in this work. Case 1 considers $f_{0}(975)$ and $\phi(1020)$ mesons while case 2 does not consider them. The unit of the cental energy density $\epsilon_{c}$ is $\times10^{15}$ g.cm$^{3}$.}
\label{tab2}
\begin{tabular}{ccccccccc}
\hline\noalign{\smallskip}
 case&$\epsilon_{c}$&$M        $&$R$   &z     &$\epsilon_{c}$&$M        $&$R$   &z         \\
    &              &M$_{\odot}$&km    &      &              &M$_{\odot}$&km    &        \\
\hline
1   &1.2470        &1.97       &12.964&0.3469&1.6721        &2.05       &12.364&0.3997    \\
2   &1.2754        &1.97       &12.941&0.3480&2.0660        &2.05       &11.907&0.4263    \\
\noalign{\smallskip}\hline
\end{tabular}
\end{center}
\end{table}

\section{Summary}
In this paper, we fit out the mass of the neutron star PSR J0348+0432 not containing or containing the mesons $f_{0}(975)$ and $\phi(1020)$ by adjusting the hyperon coupling constant in the framework of the RMF theory, respectively. Then we study the effect of the mesons $f_{0}(975)$ and $\phi(1020)$ on the radius and the surface gravitational redshift of the neutron star. We find that compared with that without considering the mesons $f_{0}(975)$ and $\phi(1020)$, the value range of the radius ( R ) will be changed from a narrow range 12.964 km $\sim$ 11.364 km to a wider range 12.941 km $\sim$ 11.907 km corresponding to the observation mass M=1.97 M$_{\odot}$$\sim$2.05 M$_{\odot}$.

For the surface gravitational redshift $z$ of the neutron star PSR J0348+0432, considering the effect of the mesons $f_{0}(975)$ and $\phi(1020)$ on the neutron star matter, its suggested value range changes from 0.3469 $\sim$ 0.3997 to 0.3480 $\sim$ 0.4263 and the central energy density $\epsilon_{c}$ changes from 1.2470$\times10^{15}$ g cm$^{-3}$ $\sim$ 1.6721$\times10^{15}$ g cm$^{-3}$ to 1.2754$\times10^{15}$ g cm$^{-3}$ $\sim$ 2.0660$\times10^{15}$ g cm$^{-3}$ corresponding to the observation mass M=1.97 M$_{\odot}$$\sim$2.05 M$_{\odot}$.

These mean the radius $R$, the surface gravitational redshift $z$ and the central energy density $\epsilon_{c}$ all will be constrained in a wider scope as the mesons $f_{0}(975)$ and $\phi(1020)$ being considered.

From the above we also see whether considering the mesons $f_{0}(975)$ and $\phi(1020)$ in the massive neutron star PSR J0348+0432 or not, the difference of the radius and the surface gravitational redshift is not so large. This indicates that the mesons $f_{0}(975)$ and $\phi(1020)$ do not play a major role in the massive neutron star PSR J0348+0432. In other words, the interactions between the nucleons and the hyperons is very weak.

\begin{acknowledgments}
We are thankful to the anonymous referee for many useful comments and suggestions and Researcher Shan-Gui Zhou for fruitful discussions.
This work was supported by the Special Funds for Theoretical Physics Research Program of the Natural Science Foundation of China ( Grant No. 11447003 ) and the Scientific Research Foundation of the Higher Education Institutions of Anhui Province, China ( Grant No. KJ2014A182 ).
\end{acknowledgments}

\bibliography{apssamp}

\end{document}